\documentclass[reprint,pra,groupedaddress,showpacs,twocolumn]{revtex4-1}
\usepackage{graphicx,bm,amsmath,amsfonts,amssymb,color}
\usepackage{multirow,color,relsize,ulem,microtype}

\newcommand{\be}{\begin{equation}}
\newcommand{\ee}{\end{equation}}
\newcommand{\bea}{\begin{eqnarray}}
\newcommand{\eea}{\end{eqnarray}}

\begin{document}
\title{Inverse Vernier Effects in Coupled Lasers}

\author{Li Ge}
\email{li.ge@csi.cuny.edu}
\affiliation{\textls[-18]{Department of Engineering Science and Physics, College of Staten Island, CUNY, Staten Island, NY 10314, USA}}
\affiliation{The Graduate Center, CUNY, New York, NY 10016, USA}
\author{Hakan E. T\"ureci}
\affiliation{Department of Electrical Engineering, Princeton University, Princeton, NJ 08544, USA}

\date{\today}

\begin{abstract}
In this report we study the Vernier effect in coupled laser systems consisting of two cavities. We show that depending on the nature of their coupling, not only can the ``supermodes" formed at the overlapping resonances of the coupled cavities have the lowest thresholds and lase first as previously suggested, leading to a manifestation of the typical Vernier effect now in an active system; these supermodes can also have increased thresholds and are hence suppressed, which can be viewed as an inverse Vernier effect. We attribute this effect to detuning-dependent $Q$-spoiling, and it can lead to an increased free spectrum range and possibly single-mode lasing, which may explain the experimental findings of several previous work. We illustrate this effect using two coupled micro-ring cavities and a micro-ring cavity coupled to a slab cavity, and we discuss its relation to the existence of exceptional points in coupled lasers.
\end{abstract}

\maketitle

\section{Introduction}
\begin{figure}[t]
\centering
\includegraphics[width=\linewidth]{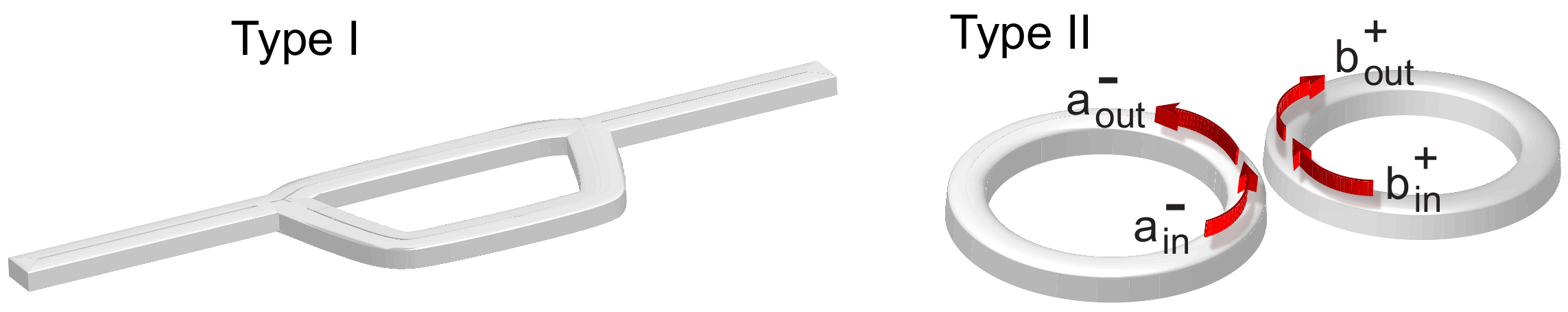}
\caption{Schematics showing two types of coupled micro-cavities. Type I utilizes an explicit interferometer setup, while Type II does not.} \label{fig:rings}
\end{figure}

The Vernier effect is well known in passive microwave and optical systems, which depicts that transmission resonances of a coupled system occur when the resonances of the subsystems coincide. It can be understood as an interference effect: destructive interference destroys all other resonances of the subsystems. The counterpart of Vernier effect in lasers has been experimentally studied with two or more coupled laser cavities, and an increased free spectral range (FSR) of the lasing spectrum and even single-mode lasing have been observed \cite{Shang_OPLETT08,Wu_APL08,Mujagic_OPEng10,Liu_APL11a,Liu_APL11b,DiDomenico,Liu_APL12,Zheng}. While some of these experiments utilized an interferometer \cite{DiDomenico,Liu_APL12,Zheng} (``Type I"; see Fig.~\ref{fig:rings}) and can be understood similar to the Vernier effect in transmission, the others were different and consisted of fused or evanescently coupled slab and micro-ring/micro-disk cavities (``Type II"). However, the understanding of the increased FSR or single-mode lasing in Type II coupled systems is still often argued using the same mechanism as in Type I systems, i.e. one cavity acts as an external cavity for frequency selection, and lasing occurs at the overlapping resonances of the coupled laser cavities.

In this report we show that frequency overlap in Type II systems does not favor lasing in general. Instead, the coupling of these overlapping resonances \textit{increases} the lowest threshold of the corresponding lasing modes. Thus the increased FSR and single-mode lasing observed can be understood as a consequence of the suppression of these overlapping modes, which is the manifestation of an \textit{inverse} Vernier effect. Below we illustrate this finding first in two evanescently coupled micro-ring cavities of different radii (see Fig.~\ref{fig:rings}) and latter in a micro-ring cavity coupled to a slab cavity. We show that the changes of the lasing thresholds are related to the existence of exceptional points (EPs) \cite{EP1,EP2,EPMVB,EP3,EP4,EP5,EP6,EP7,EP_CMT}, at which two lasing modes have the same frequency, threshold, \textit{and} spatial intensity pattern.
% and the much weaker dependence of the detuned resonances on coupling.
We further show that the effect of coupling in Type I systems increases with the detuning between two neighboring resonances, one in each of the two coupled cavities, while that in Type II systems \textit{decreases} with the detuning, yielding the inverse Vernier effect instead of the typical Vernier effect. Finally, we reveal that 
%while the typical Vernier effect does not rely on the optical losses of the constitute cavities, 
the inverse Vernier effect depends on different optical losses of the coupled cavities, or equivalently, their different quality ($Q$) factors, highlighting its origin in coupling-caused $Q$-spoiling.

Our analysis is based on the coupled-mode formulism suggested by Yariv \cite{Yariv}, which takes into account the amplitude and phase evolution of light inside the coupled cavities. Since the increased FSR and single-mode lasing reported in Refs.~\cite{Shang_OPLETT08,Wu_APL08,Mujagic_OPEng10,Liu_APL11a,Liu_APL11b} were observed close to the lowest lasing threshold, nonlinearity was not crucial for these observation and we neglect it in the analysis below. We first consider two coupled micro-ring cavities (see Fig.~\ref{fig:rings}), and the coupling between them can be captured by a scattering ($S$) matrix \cite{Yariv}:
\be
\begin{pmatrix}
a^-_\text{out}\\
b^+_\text{out}
\end{pmatrix} =
S
\begin{pmatrix}
a^-_\text{in}\\
b^+_\text{in}
\end{pmatrix},\quad
S =
\begin{pmatrix}
t & J \\
-J^* & t^*
\end{pmatrix},\label{eq:S}
\ee
where $a^-_\text{in,\,out}$ are the incoming and outgoing counterclockwise wave to the coupling junction in the first cavity, and $b^+_\text{in,\,out}$ are similar defined for the clockwise waves in the second cavity. The coupling of the waves traveling in the opposite directions, i.e. $a^+_\text{in,\,out}$ and $b^-_\text{in,\,out}$, is given by the same $S$ matrix because of the local spatial symmetry at the coupling junction. The $S$ matrix is unitary and conserves the local flux, i.e., $|t|^2 + |J|^2=1$. Since we do not expect a phase jump when $a^-_\text{in}$ passes through the coupling junction to become part of $a^-_\text{out}$, we take $t$ to be real.

Assuming the circumferences of the two ring cavities are $L_1$ and $L_2$, the phase and amplitude changes of light after one circulation in each cavity and before coupling again is given by
\begin{align}
a^{-}_\text{in} &= e^{i(n+i\kappa_1-i\tau)kL_1}\, a^{-}_\text{out} \equiv \beta_1 a^{-}_\text{out},\label{eq:a}\\
b^{+}_\text{in} &= e^{i(n+i\kappa_2-i\tau)kL_2}\, b^{+}_\text{out} \equiv \beta_2 b^{+}_\text{out},\label{eq:b}
\end{align}
respectively. Here $n$ is the refractive index of the ring cavities and $k=\omega/c$ is the wave number in free space. For simplicity, we take $c=1$ and do not distinguish between $k$ and the frequency $\omega$. The optical losses (including radiation loss, material absorption, etc.) are represented by $\kappa_1,\kappa_2$ in these two cavities, respectively, and to focus on coupling-induced threshold changes we will treat them as constants for all modes. The optical gain is modeled by adding a negative imaginary part $-i\tau$ to the refractive index \cite{linearGain1,linearGain2}.

By solving Eqs.~(\ref{eq:S}) and (\ref{eq:b}), we find the following relation between the two counterclockwise amplitudes in the first micro-ring cavity:
\be
a_\text{out}^- = \frac{t-\beta_2}{1-t\beta_2}\,a_\text{in}^-.\label{eq:aout}
\ee
%from which the well-known critical coupling condition $t=\beta_2\neq1$ for a vanished $a_\text{out}^-$ is readily seen.
The lasing thresholds are determined by the self-consistent condition imposed by Eqs.~(\ref{eq:a}) and (\ref{eq:aout}), e.g., $a_\text{in}^-$ should not change in steady-state lasing oscillation after light circulates the first ring cavity once and comes back to the same location:
\be
\beta_1\frac{t-\beta_2}{1-t\beta_2} = 1.\label{eq:roundtrip}
\ee

In the absence of coupling, i.e. $J=0$ and $t=1$, we recover the simple relation $\beta_1 = 1$ that determines the lasing frequencies and thresholds of the first micro-ring cavity, i.e.
\be
k_{1,m} = \frac{2\pi m}{n L_1},\; \tau_{1,m} = \kappa_1 \;(m=1,2,\ldots)\nonumber
\ee
Similarly, the lasing modes in the second micro-ring cavity are given by $k_{2,m} = {2\pi m}/nL_2$ and $\tau_{2,m}=\kappa_2$. In order to recover these results for the second micro-ring cavity in the absence of coupling, i.e. $\beta_2=1$, it is necessary to rewrite Eq.~(\ref{eq:roundtrip}) in the following equivalent form:
\be
\beta_2\frac{t-\beta_1}{1-t\beta_1} = 1. \label{eq:roundtrip2}
\ee

In the opposite limit of strong coupling, i.e. $|J|\rightarrow1$ and $t\rightarrow0$, both Eqs.~(\ref{eq:roundtrip}) and (\ref{eq:roundtrip2}) become
\be
\beta_1\beta_2=-1, \label{eq:roundtrip3}
\ee
which indicates that the system is now effectively a micro-ring cavity of circumference ($L_1+L_2$).
% with a $\pi$-phase shifter (coming from the ``$-$" sign) due to the coupling. 
We note that this result, as well as Eqs.~(\ref{eq:roundtrip}) and (\ref{eq:roundtrip2}), do not depend on the phase of the coupling $J$. Therefore, we take $J$ to be real in the following discussions.

\section{Inverse Vernier Effect}

\begin{figure}[b]
\centering
\includegraphics[width=\linewidth]{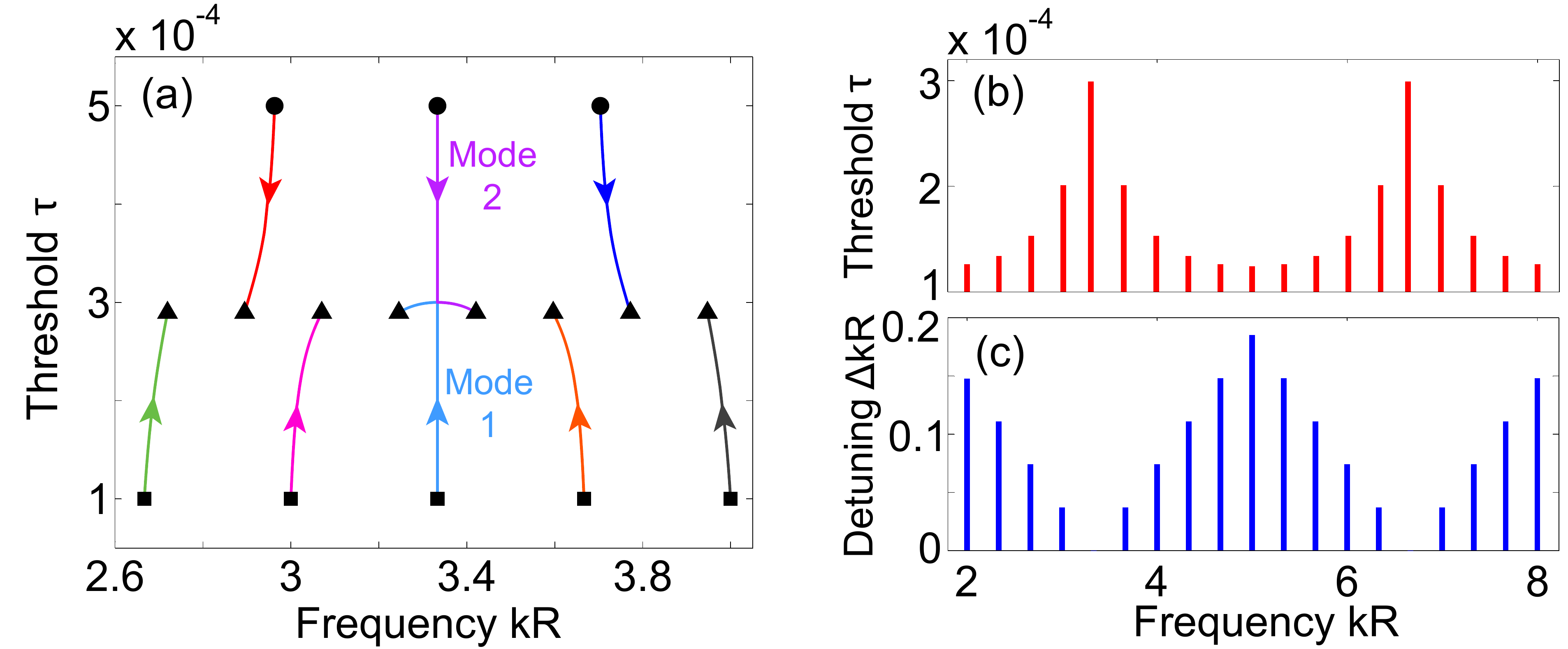}
\caption{(Color Online) Inverse Vernier effect in two evanescently coupled micro-ring cavities. (a) Trajectories of the lasing thresholds versus the frequencies as the coupling $J$ increases from 0 to 1. The squares and dots mark the uncoupled lasing modes at $J=0$, respectively. The triangles show the coupled lasing modes at $J=1$. Arrows indicate the direction of motion as $J$ increases. $R$ and $0.9R$ are the radius of the larger and smaller cavities, respectively. The total loss in the two cavities are $\kappa_1=10^{-4}$ and $\kappa_2=5\times10^{-4}$, respectively. The refractive index is $n=3$. (b) Lowest thresholds of the lasing modes at $J=0.5$, which evolve from the uncoupled resonances in the larger ring cavity. Note their increased thresholds from $\kappa_1$, especially at the perfectly aligned resonances near $kR=3.3,6.6$. (c) shows the detuning of these uncoupled resonances with the nearest counterparts in the smaller ring cavity.} \label{fig:largeCoupling}
\end{figure}

The FSRs of the uncoupled micro-ring cavities are $\Delta k_1 = 2\pi/nL_1$ and $\Delta k_2 = 2\pi/nL_2$, respectively. The average spectral density is then given by $\Delta k_1^{-1} + \Delta k_2^{-1}$, not counting the double degeneracy of the micro-ring resonances due to the clockwise and counterclockwise symmetry. Note that it is equal to the spectral density given by Eq.~(\ref{eq:roundtrip3}) at $J=1$, where the lasing frequencies and thresholds are given by
\begin{gather}
k_m = \frac{(2 m+1)\pi}{n(L_1+L_2)},\label{eq:roundtrip3_k}\\
\tau_m = \frac{L_1 \,\kappa_1   + L_2 \kappa_2}{L_1 + L_2}. \label{eq:roundtrip3_tau}
\end{gather}
This observation indicates that the lasing modes in the coupled system evolve continuously from the uncoupled resonances as $J$ increases from 0 to 1 [see Fig.~\ref{fig:largeCoupling}(a)], with the ones originating from the larger ring cavity having the lower thresholds.
The thresholds (\ref{eq:roundtrip3_tau}) at $J=1$ are the same for all modes, given by the average of the thresholds of the uncoupled micro-ring cavities and weighted by the corresponding circumference.

We note, however, that this observation does not mean that the thresholds of the lasing modes have the same dependence on the coupling. As can be seen from Fig.~\ref{fig:largeCoupling}(b) at $J=0.5$, there is a clear difference between the thresholds of the lasing modes, which are inversely correlated with the detuning of the uncoupled resonances. The least overlapped resonances of the larger ring cavity have the lowest threshold and lase at a low pump power, while the better overlapped ones have higher thresholds and are suppressed at a low pump power. We refer to this effect as the \textit{inverse} Vernier effect, since it is in opposite to the Vernier effect in transmission that preserves only the overlapping resonances. Nevertheless, the FSR of the active lasing modes can also be increased as a result, and single-mode lasing may become possible if the gain spectrum is not too wide.

\begin{figure}[b]
\centering
\includegraphics[width=\linewidth]{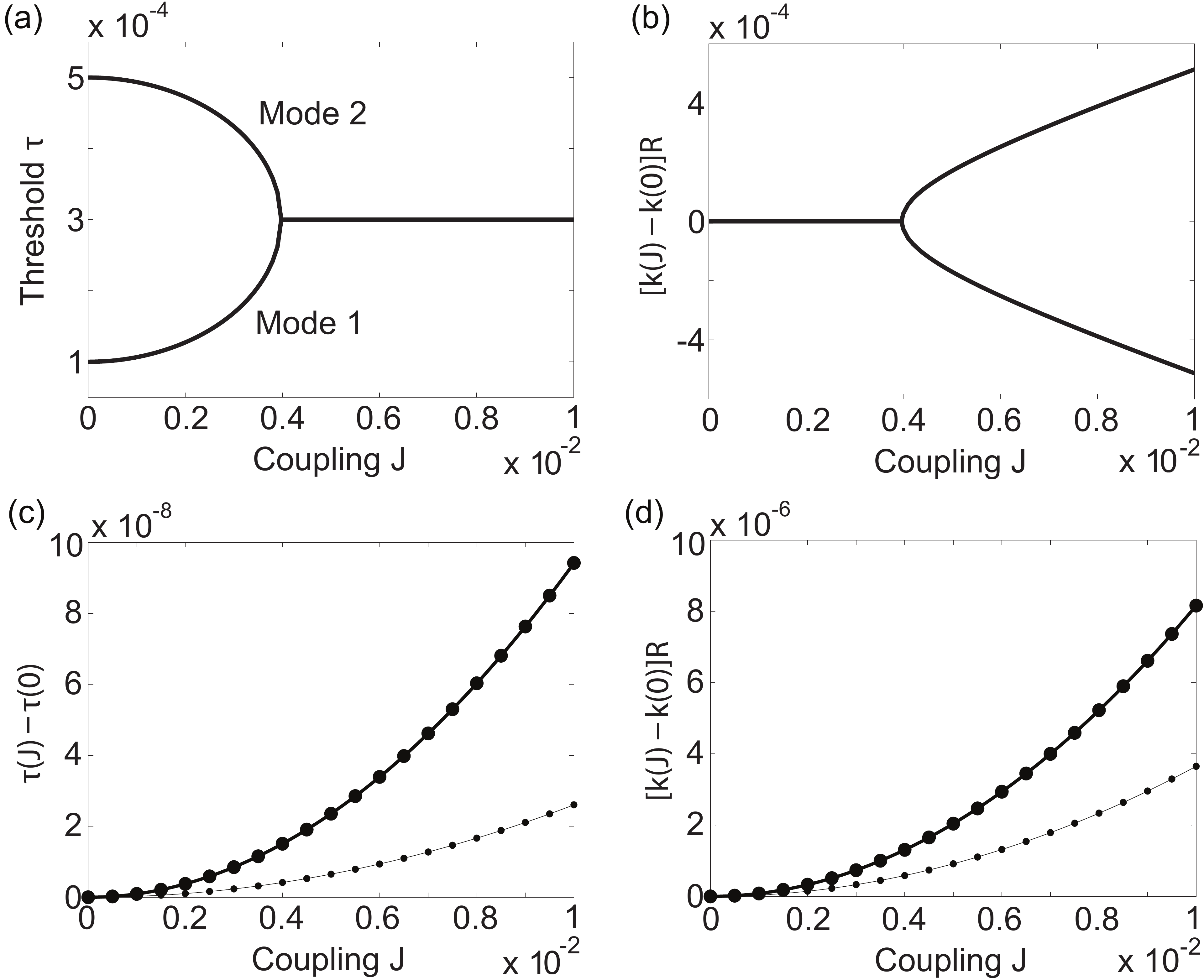}
\caption{Inverse bifurcation of the lasing frequencies (a) and bifurcation of the corresponding thresholds (b) of the perfectly aligned mode 1 and 2 near $kR=3.3$ shown in Fig.~\ref{fig:largeCoupling}(a), as the coupling increases from $J=0$.
The bifurcations occur near $J=4\times10^{-3}$. The solid lines in (c) and (d) show the much weaker $J$-dependencies of the two low-threshold modes on the left of mode 1 in Fig.~\ref{fig:largeCoupling}(a), with the thin one further away from mode 1. The dots show the analytical approximations given by Eqs.~(\ref{eq:dk}) and (\ref{eq:dtau}).} \label{fig:bifurcation}
\end{figure}

To better understand the much stronger $J$-dependence of the thresholds at the spectrally aligned resonances (e.g., mode 1 of the larger cavity and mode 2 of the smaller cavity in Fig.~\ref{fig:largeCoupling}(a)), we first note one of their qualitative differences from the detuned resonances. Starting from $J=0$, mode 1 and 2 first move vertically in the $k$-$\tau$ plane with an increasing coupling and coalesce before moving sideways. The detuned resonances, on the other hand, undergo avoided crossings instead.
This behavior of mode 1 and 2 is plotted as a function of the coupling $J$ in Fig.~\ref{fig:bifurcation}(a) and (b). Their frequencies and thresholds experience a bifurcation and reverse bifurcation respectively when $J$ becomes $J_c\approx4\times10^{-3}$, which indicate the existence of an EP \cite{EP1,EP2,EPMVB,EP3,EP4,EP5,EP6,EP7,EP_CMT}. In contrast, the detuned resonances show a much weaker $J$-dependence when $J$ is small, as shown in Fig.~\ref{fig:bifurcation}(c) and (d): At $J=J_c$ the threshold increase of mode 1 is more than $10^4$ times larger than the detuned resonances.
%This weak quadratic dependence further reduces as the detuning of two uncoupled resonances, one from each cavity, increases. This behavior will be investigated analytically in the next section.

\begin{figure}[t]
\centering
\includegraphics[width=\linewidth]{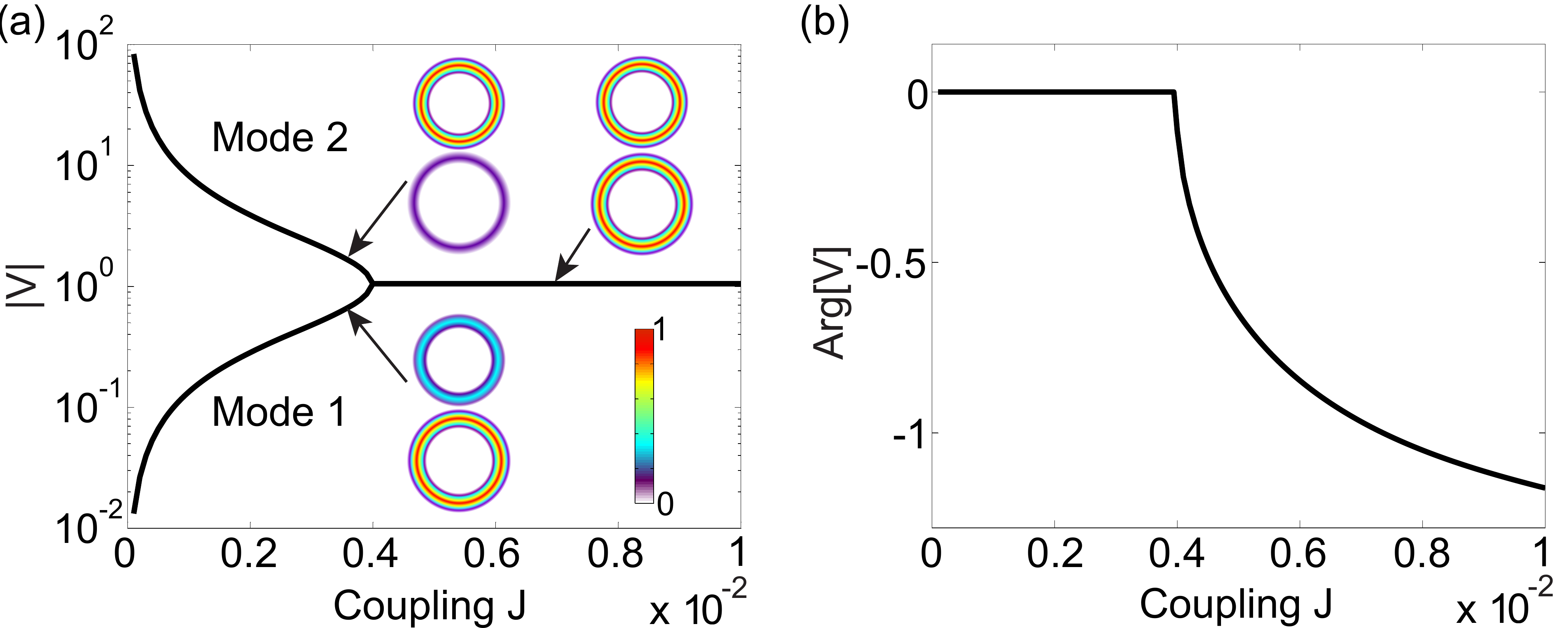}
\caption{(Color online) The ratio $V$ of light amplitudes inside the two micro-ring cavities for mode 1 and 2 in Fig.~\ref{fig:bifurcation}(a) and (b). $V$ is defined in Eqs.~(\ref{eq:ratio1}) and (\ref{eq:ratio2}). The insets in (a) illustrate their different intensity ratios at $J=3.5\times10^{-3}$ below the EP at $J_c=4\times10^{-3}$ and their identical intensity ratio $|V|=1$ at $J=7\times10^{-3}$ above the EP. (b) The identical phase of $V$ for mode 1 and 2, both below and beyond $J_c$.} \label{fig:ratio}
\end{figure}

The EPs are often studied in an eigenvalue problem \cite{EP1,EP2,EPMVB}. Although in our coupled-mode formulism the threshold conditions (\ref{eq:roundtrip}) and (\ref{eq:roundtrip2}) do not have the explicit form of an eigenvalue problem, the merging of the frequencies and thresholds of mode 1 and 2 shown in Figs.~\ref{fig:largeCoupling}(a) and \ref{fig:bifurcation}(a)(b) at $J=J_c$ is a clear indication of an EP. This is further confirmed by the coalescence of their wavefunctions (see Fig.~\ref{fig:ratio}), which distinguishes an EP from a usual degeneracy point. In the next section we will analyze the location of the EP as well as the much weaker $J$-dependence of the detuned resonances.

\section{Analytical results and physical interpretations}

When the coupling is small, mode 1 and 2 concentrate in the larger and smaller micro-rings, respectively. This can be seen quantitatively from
\be
V\equiv\frac{b^+_\text{in}}{a^+_\text{in}} = \frac{-\beta_2 J}{1-t\beta_2}\label{eq:ratio1}
\ee
for mode 1: the lasing condition of the first micro-ring cavity, i.e. $\beta_1=1$, holds for this mode when $J\rightarrow0$, and $\beta_2\neq1$ because of the different thresholds of mode 1 and 2 when they are uncoupled. Therefore, $V\rightarrow0$ for mode 1, which has little amplitude in the second micro-ring cavity as expected.

Similarly, $V$ can be expressed as
\be
V = \frac{1-t\beta_1}{\beta_1 J}\label{eq:ratio2}
\ee
for mode 2: the lasing condition of the second micro-ring cavity, i.e. $\beta_2=1$, holds for this mode when $J\rightarrow0$, and $\beta_1\neq1$ because of the different thresholds of mode 1 and 2 when they are uncoupled. Therefore, $V\rightarrow\infty$ for mode 2, which has little amplitude in the first micro-ring cavity as expected.

We note that the two expressions (\ref{eq:ratio1}) and (\ref{eq:ratio2}) are mathematically identical using Eq.~(\ref{eq:roundtrip}) or (\ref{eq:roundtrip2}). We discussed them separately above just to avoid the ratio of two vanishing quantities in the limit $J\rightarrow0$. Once $J$ becomes finite, either expression can be used for both mode 1 and 2, and their $V$ values (and hence their wavefunctions) become the same once they have the same value of $\beta_1$ (and consequently $\beta_2$ as well). This condition is satisfied when the frequencies and thresholds of these two modes become the same, i.e., at the EP.

To locate the EP in terms the coupling $J$, we first note that at the aligned resonant frequency $k=k_0$, both $\beta_1=\exp[(\tau-\kappa_1)k_0L_1]\equiv\tilde{\beta}_{1}$ and $\beta_2=\exp[(\tau-\kappa_2)k_0L_2]\equiv\tilde{\beta}_{2}$ are real-valued. Consequently, Eq.~(\ref{eq:roundtrip}) can be solved at $k=k_0$, with the threshold $\tau$ determined by
\be
t = \frac{1+\tilde{\beta_1}\tilde{\beta_2}}{\tilde{\beta_1}+\tilde{\beta_2}}. \label{eq:EP}
\ee
For cavities of relatively high quality factors (and hence with low losses), the exponents in $\tilde{\beta}_{1,2}$ are very small and we expand them to the second order of $\tau$, which gives rise to
\be
(\tau-\kappa_1)(\tau-\kappa_2)\approx\frac{2(t-1)}{k_0^2L_1L_2}.\label{eq:EP2}
\ee

The left hand side depicts a quadratic curve of $\tau$, with the minimum $(\kappa_1-\kappa_2)^2/4$ at $\tau=(\kappa_1+\kappa_2)/2$. If this minimum is lower than the constant on the right hand side, i.e.
\be
t>1-\frac{k_0^2L_1L_2}{8}(\kappa_1-\kappa_2)^2,
\ee
or equivalently,
\be
J<J_c\equiv\frac{1}{2}k_0\sqrt{L_1L_2}|\kappa_1-\kappa_2|,\label{eq:Jc}
\ee
Eq.~(\ref{eq:EP2}) gives two real solutions of $\tau$ (i.e., mode 1 and 2).
Right at $J=J_c$, these two solutions coalesce into one, and the EP is reached. If $J$ is larger than $J_c$, then there is no solution to Eq.~(\ref{eq:EP}) with a real $\tau$, which means that the corresponding modes can no longer exist at $k=k_0$, leading to the frequency bifurcation shown in Fig.~\ref{fig:bifurcation}(b).

Equation (\ref{eq:Jc}) gives $J_c=3.97\times10^{-3}$ for the example shown in Fig.~\ref{fig:bifurcation}, which agrees well with the numerical result for the location of the bifurcations shown. Equation~(\ref{eq:Jc}) also shows that the toy model given in Ref.~\cite{EP_CMT} is qualitatively correct, and the location of an EP in terms of the coupling is proportional to the difference of the losses in the two coupled cavities.

Similar to the derivation above, we obtain the approximations for the frequency and threshold changes of the lower-threshold modes, originating from the uncoupled resonances in the larger ring cavity:
\begin{align}
\delta k(J)&\equiv k(J)-k(0)
\approx \frac{J^2}{2nL_1}\frac{\sin\theta}{1-\cos\theta},\label{eq:dk}\\
%\approx \frac{J^2}{4\pi^2n^2RR'\Delta}, \label{eq:dk}\\
\delta \tau(J)&\equiv \tau(J)-\tau(0)
%\approx \frac{(\kappa_2-\kappa_1)\delta k(J)}{\Delta},
\approx \frac{(\kappa_2-\kappa_1)\delta k(J)}{\Delta}\frac{\theta}{\sin\theta}.
\label{eq:dtau}
\end{align}
Here $\theta\equiv n\Delta L_2$ and $\Delta$ is the detuning of these resonant frequencies from the nearest ones in the smaller cavity.
Equations~(\ref{eq:dk}) and (\ref{eq:dtau}) give excellent agreement with the numerical results when $J$ is small, as can be seen from Fig.~\ref{fig:bifurcation}(c) and (d). They show that both $\delta k$ and $\delta \tau$ are proportional to $J^2$ when $J$ is small, and more importantly, these changes are inversely correlated with the detuning $\Delta$ when $|\theta|\ll1$, with $\delta k(J)$ proportional to $\Delta^{-1}$ and $\delta \tau(J)$ proportional to $\Delta^{-2}$ in this limit.
%and the difference of the losses in the two micro-ring cavities.
Due to the two different FSRs of the two coupled cavities, the detuning $\Delta$ modulates as a function of frequency and so does the lasing threshold $\tau$, which then leads to the inverse Vernier effect of the lasing modes when the pump power is low.

This finding can be interpreted in the following way: for two cavities of different losses (and hence different $Q$-factors), the coupling effect is strong for overlapping resonances, and the higher-$Q$ resonances are ``spoiled" more strongly by the lower-$Q$ ones, causing a significant increase of their thresholds. For little- or non-overlapping resonances, this $Q$-spoiling effect is weak, and hence the thresholds of the higher-$Q$ resonances do not vary much from their uncoupled values.

From this interpretation it is clear that different $Q$ factors in the two coupled cavities is crucial for the inverse Vernier effect, which would not occur if the losses in the two micro-ring cavities are the same; this can be directly seen from Eq.~(\ref{eq:dtau}), which shows that the lasing threshold $\tau$ does not change with the detuning $\Delta$ if $\kappa_1=\kappa_2$. A more rigorous proof without using the expansion for a small coupling $J$ is given in the appendix.

As we discussed above, the inverse Vernier effect in Type II coupled systems is the result of detuning-dependent $Q$-spoiling due to the coupling to a lower-$Q$ cavity. The typical Vernier effect, on the other hand, is caused by the detuning-dependent destructive interference. To contrast their different dependencies on the detuning $\Delta$, below we use the Michelson interferometer setup \cite{DiDomenico} to exemplify Type I systems, the threshold condition of which is given by
\be
\beta_{1}(k,\tau)T + \beta_{2}(k,\tau)R=1, \label{eq:roundtrip_michelson}
\ee
where $\beta_{1,2}=e^{i(n+i\kappa_{1,2}-i\tau)kL_{1,2}}$ are the phase and amplitude changes after one circulation along each arm of the Michelson interferometer. $T$ and $R=1-T$ are the transmittance and reflectance of the beam splitter, and $L_{1,2},\kappa_{1,2}$ are the length and loss of each arm. For $T=0$ or $1$, lasing in the two arms takes place independently.

For simplicity, we consider a 50/50 beam splitter ($T=R=0.5$), which simplifies the threshold condition to $\beta_{1}(k,\tau) + \beta_{2}(k,\tau)=2$. Similar to the derivation for the Type II coupled systems, we find the threshold change of the higher-$Q$ modes is given by
\be
\delta\tau = \tau(T=0.5) - \tau(T=0) \approx \frac{\cos\theta-1-\frac{L_2\sin^2\theta}{L_1+L_2\cos\theta}}{k_0L_1}. \label{eq:dtau_typeI}
\ee
We note that Eq.~(\ref{eq:dtau_typeI}) is proportional to the detuning $\Delta^2$ when $|\theta|\ll1$. In other words, the effect of coupling, or more precisely, the effect of destructive interference, is more pronounced for a larger detuning as expected. This is in stark contrast with the relation (\ref{eq:dtau}) for the threshold change in Type II coupled systems ($\delta\tau\propto \Delta^{-2}$), which distinguishes the typical Vernier effect and the inverse Vernier effect reported here.

We also note that the difference of the losses, $\kappa_1-\kappa_2$, does not appear in Eq.~(\ref{eq:dtau_typeI}); it is a higher order term for high-$Q$ modes, or more specifically, when $|\kappa_1-\kappa_2|{k_1L_2}\ll1$. Thus the typical Vernier effect in Type I systems is not related to $Q$-spoiling due to the coupling to a lower-$Q$ cavity, while this mechanism is what causes the inverse Vernier effect in Type II coupled systems as discussed.

\section{Discussion and Conclusion}

\begin{figure}[b]
\centering
\includegraphics[width=0.8\linewidth]{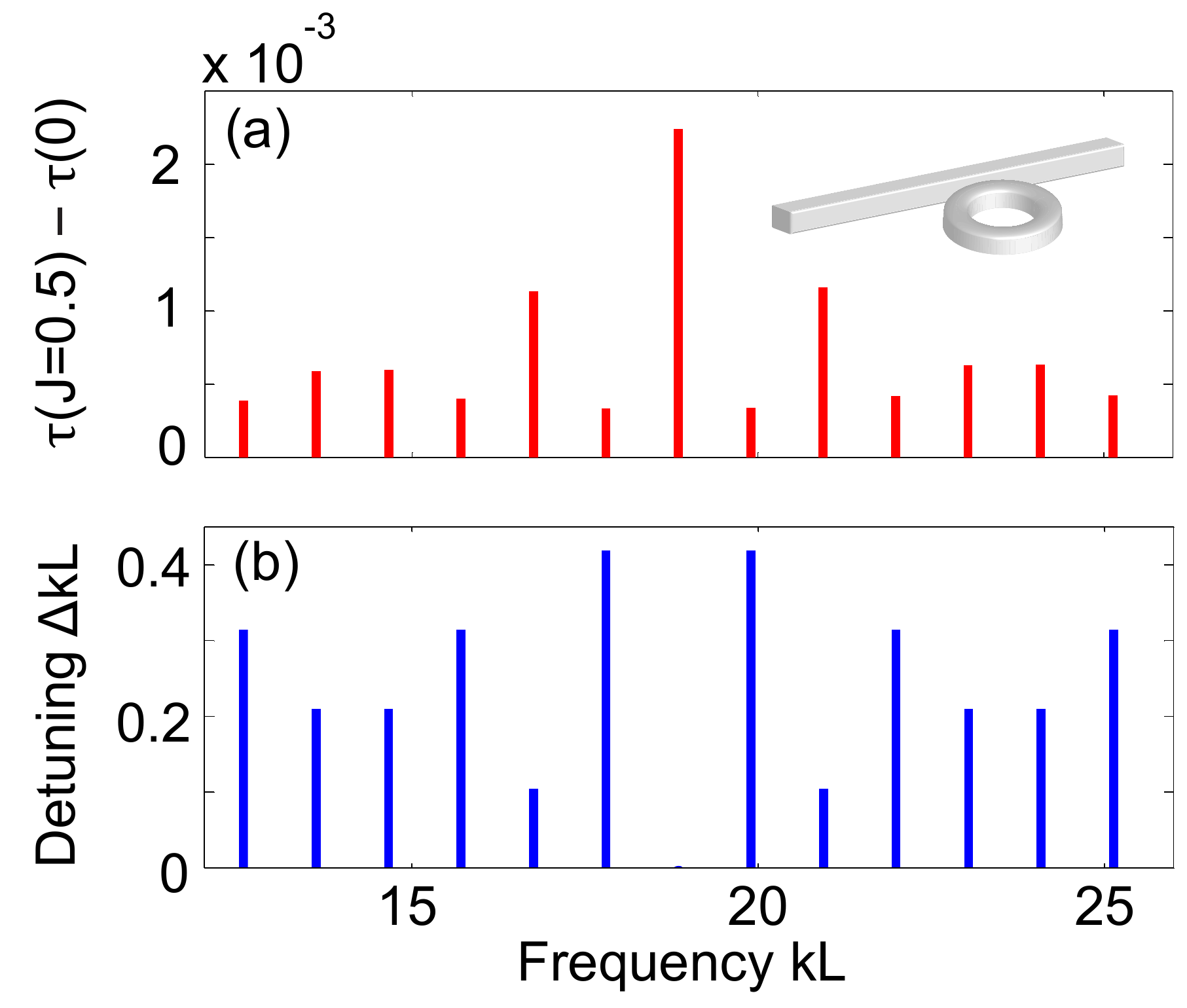}
\caption{(Color online) Inverse Vernier effect in a slab cavity of length $L$ coupled with a micro-ring cavity of radius $R=L/1.8\pi$ [see the inset in (a)]. (a) Threshold changes for the lowest threshold modes near $kL=20$ at $J=0.5$. They originate from the uncoupled slab resonances, the loss of which is %$\kappa_S\approx5.0\times10^{-4}$ 
assumed to come from the radiation through two facet mirrors of reflectivity $r=0.99$ and lower than that in the micro-ring ($\kappa=5\times10^{-3}$). (b) Their threshold change is inversely correlated with their detuning from the nearest micro-ring resonance.
} \label{fig:bar_ring}
\end{figure}

Our analysis based on the coupled-mode formulism is general and can also be applied to, for example, a slab cavity coupled with a micro-ring/micro-disk cavity. The only differences are: (i) a different $\beta$ factor is needed to capture the phase and amplitude change of the light after a round trip in the slab cavity, including the effect of the radiation loss through the end facets; and (ii) the clockwise and counterclockwise waves in the micro-ring/micro-disk cavity are coupled by a slab resonance. More specifically, the equivalence of the threshold condition (\ref{eq:roundtrip}) or (\ref{eq:roundtrip2}) is
\be
\beta_r\frac{t-\beta_s}{1-t\beta_s} = 1,\label{eq:roundtrip_barRing}
\ee
where $\beta_r$ is defined in the micro-ring cavity similar to $\beta_1$ in Eq.~(\ref{eq:a}) and $\beta_s\equiv\pm e^{i(n+i\kappa_s-i\tau)kL}$ is defined in the slab cavity of length $L$ and loss $\kappa_s$. If the radiation loss from the two facets of reflectivity $r$ dominates the losses in the slab cavity, $\kappa_s$ is then given approximately by $-\ln(r)/2kL$. The inverse Vernier effect still holds, as we show in Fig.~\ref{fig:bar_ring}.

In summary, we have shown that for two coupled cavities of different FSRs, the overlapping of their resonances do not typically favor lasing, %unless in an interferometer setup, 
resulting in an inverse Vernier effect. The suppression of these overlapping resonances can also lead to an increased FSR and possibly single-mode lasing, as found in previously experiments \cite{Shang_OPLETT08,Wu_APL08,Mujagic_OPEng10,Liu_APL11a,Liu_APL11b}. We have treated the coupling $J$ in our model (\ref{eq:S}) as a constant for all modes. If we consider a weaker value of the coupling due to a larger detuning, the differences of the maximum and minimum thresholds shown in Figs.~\ref{fig:largeCoupling}(b) and \ref{fig:bar_ring}(a) will be smaller, but their qualitative modulation as a function of the frequency still holds, and hence so does the inverse Vernier effect.
%For the resonances with a significant detuning, i.e. comparable with their passive linewidths $\Gamma$, their coupling should be lower due to the detuning.

We thank Peter Liu, Ramy El-Ganainy and Claire Gmachl for helpful discussions. This project was partially supported by PSC-CUNY 45 Research Grant, NSF Grant No. MIRTHE EEC-0540832, and DARPA Grant No. N66001-11-1-4162.

\appendix
\section*{Appendix: Role of different cavity losses}
In the main text we discussed that the different losses in coupled cavities is the key factor that leads to the inverse Vernier effect. This observation was based on the physical interpretation and the expansion of the threshold condition in the weak coupling limit ($J\ll1$). Here we show more rigorously that the lasing threshold $\tau$ does not change with the coupling $J$ if the coupled cavities have the same loss, i.e., $\kappa_1=\kappa_2$, and hence the inverse Vernier effect does not occur in this case.

\begin{figure}[t]
\centering
\includegraphics[width=0.7\linewidth]{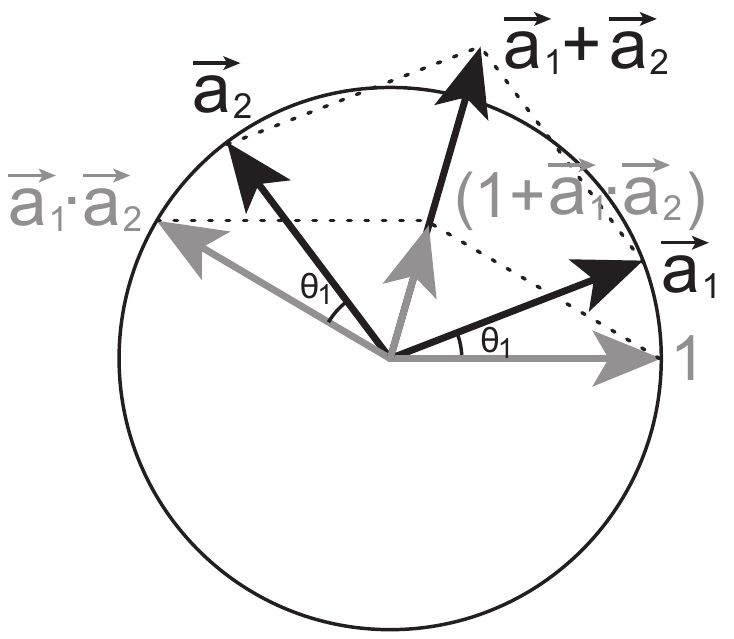}
\caption{Schematics showing the right hand side of Eq.~(\ref{eq:fixedtau}).} \label{fig:fixedtau}
\end{figure}

What we do is the following: we take $\tau$ to be equal to $\kappa_1=\kappa_2$, and show that the resulting threshold condition
\be
t \in[0,1] = \frac{1+e^{ink(\theta_1+\theta_2)}}{e^{i\theta_1}+e^{i\theta_2}} \label{eq:fixedtau}
\ee
can be satisfied simply by varying the lasing frequency $k$. Here $\theta_1\equiv nkL_1$ and $\theta_2\equiv nkL_2$ are the phase changes in the two ring cavities after a round trip. We note that the right hand side of Eq.~(\ref{eq:fixedtau}) depicts the sum of two unit vectors $\vec{a_1},\vec{a_2}$ dividing the sum of their inner product ($\vec{a_1}\cdot\vec{a_2}$) and the unit vector along the real axis. From the phasor diagram shown in Fig.~\ref{fig:fixedtau}, we know that these two sums are both along the bisector of the angle formed by $\vec{a_1}$ and $\vec{a_2}$, because $1$ is rotated from $\vec{a_1}$ clockwise by $\theta_1$ and $\vec{a_1}\cdot\vec{a_2}$ is rotated from $\vec{a_2}$ counterclockwise by the same angle. Therefore, their ratio is indeed real as required by Eq.~(\ref{eq:fixedtau}). Eq.~(\ref{eq:fixedtau}) at any coupling $J=\sqrt{1-t^2}$ can then be satisfied by varying $\theta_1$ and $\theta_2$ via $k$, which changes the ratio of the two aforementioned sums. This concludes our proof.

%To further confirm, without using any approximation, that the threshold does not change at all when $\kappa_1=\kappa_2$, we consider a simpler example, with two identical micro-ring cavities (and hence $\beta_1=\beta_2\equiv\beta$), and the threshold conditions (\ref{eq:roundtrip}) and (\ref{eq:roundtrip2}) become
%\be
%1+\beta^2 = 2t\beta,
%\ee
%the solutions of which are given by $\beta=(t\pm i|J|)^{-1}$. Since $|t|^2+|J|^2=1$ from the current conservation at the coupling junction and we have taken $t$ to be real, we find that $|\beta|=1$ always holds in this case, or equivalently, the threshold does not change its value $\tau=\kappa_1=\kappa_2$) when the coupling $J$ varies.

\bibliographystyle{apsrev}

\end{document}